\documentclass{aa}
\usepackage{graphics}

\def\bc{\begin{center}}
\def\ec{\end{center}}
\def\be{\begin{equation}}
\def\ee{\end{equation}}
\def\bea{\begin{eqnarray}}
\def\eea{\end{eqnarray}}

\begin{document}
\title{Database of optical properties of cosmic dust analogues (DOP)}
\titlerunning{Database of optical properties}

\author{
         V.B.~Il'in\inst{1,2},
         N.V.~Voshchinnikov\inst{1,2},
         V.A.~Babenko\inst{3},
         S.A.~Beletsky\inst{4},
         Th.~Henning\inst{5},
         C.~Jager\inst{6},
         N.G.~Khlebtsov\inst{7,8},
         P.V.~Litvinov\inst{9},
         H.~Mutschke\inst{6},
         V.P.~Tishkovets\inst{4}, and
         R.~Waters\inst{10}
         }
\authorrunning{Il'in et al.}

\institute{
 Sobolev Astronomical Institute,
 St.~Petersburg State University,
 Russia
\and
 St.~Petersburg Branch of Isaac Newton Institute of Chile
\and
 Stepanov Institute of Physics, Minsk, Belarus
\and
 Astronomical Observatory, Kharkov National University, Ukraine
\and
 Max Planck Institute for Astronomy, Heidelberg, Germany
\and
 Astrophysical Institute and University Observatory,
 Friedrich Schiller University, Jena, Germany
\and
 Institute of Biochemistry and Physiology of Plants and Microorganisms,
 Saratov, Russia
\and
 Saratov State University, Russia
\and
 Institute of Radio Astronomy, Kharkov, Ukraine
\and
 Astronomical Institute, University of Amsterdam, The Netherlands
}

\offprints{V.B. Il'in, vi2087@vi2087.spb.edu}

\date{Received $<$date$>$; accepted $<$date$>$}

\abstract{We present a database containing information
on different aspects of calculation and usage of the optical
properties of small {\it non-spherical} particles --- cosmic dust analogues.
 The main parts of the DOP are
a review of available methods of the light scattering theory,
collection of light scattering codes, special computational tools,
a graphic library of the optical properties
(efficiencies, albedo, asymmetry factors, etc.),
a database of the optical constants for astronomy,
a bibliographic database of light scattering works,
and links to related Internet resources.
 The general purpose of the DOP
having the address {\it http://www.astro.spbu.ru/DOP} is twofold ---
to help scientists to apply the light scattering theory in astronomy
and to give students and beginners
a possibility quickly to get necessary knowledge on the subject.
\keywords{databases, cosmic dust, light scattering}
      }
\maketitle

\section{Introduction}

Dust has been detected in most astronomical objects ---
from the Solar system to active galactic nuclei, and everywhere
the dust grains were found to play an important role
in different physical and chemical processes.
Modelling of these objects should
include accurate consideration of scattering, absorption
and emission of radiation by the dust grains.

So far in such a modelling one mainly utilized the model
of homogeneous spherical particles, though there are many
evidences that the interstellar grains are mostly non-spherical
and inhomogeneous.
 Today enough possibilities exist to avoid the use
of the spherical model.
 To help scientists to make this step, we have developed
a database including useful information, data, codes,
references and links
and made it accessible via the Internet.
 As far as we know there are no analogous databases
in the World Wide Web.

In this paper our database
having the address {\it http://www.astro.spbu.ru/DOP}
is described first briefly, and then partly in more detail.
 The database contains various introductory notes written for
students and beginners and more specific information on light scattering
theory and its applications.
 The DOP includes many pieces of the review ``Optics of cosmic dust''
by Voshchinnikov~(\cite{Vo02}) and materials being used in lectures on
physics of interstellar matter in the St.~Petersburg University.

\section{General overview of the database}

The information, resources, and links presented in the DOP
are collected in three main blocks:

1) those related with the theoretical aspects~---
   scatterer models, methods of their realization, bibliography;

2) those connected with calculations (and hence applications)~---
     optical constants, light scattering codes,
     graphic and tabular libraries of the optical properties, special tools;

3) those coupled with other topics (e.g., radiative transfer modelling).


The database has the following sections and subsections:
\newpage
\begin{itemize}
\item[A.] {\bf Theoretical aspects}
 \begin{itemize}
 \item[A1.] Definitions of the optical characteristics of non-spherical scatterers
 \item[A2.] Scatterer models and their parameters
 \item[A3.] Exact and approximate methods of light scattering theory
 \item[A4.] Light scattering experiments
 \item[A5.] Bibliography of light scattering works
 \end{itemize}
\item[B.] {\bf Application aspects}
 \begin{itemize}
 \item[B1.] Optical constants of materials
 \item[B2.] Light scattering codes
 \item[B3.] Benchmark results
 \item[B4.] A graphic library of the optical properties
 \item[B5.] Self-training algorithm for calculation of the optical
       properties of fractal-like aggregates
 \end{itemize}
\item[C.] {\bf Related resources}
 \begin{itemize}
 \item[C1.] Radiative transfer tools
 \item[C2.] Miscellaneous
 \end{itemize}
\end{itemize}
Let us now turn to a brief description of the subsections.

\subsection{DOP subsections on theoretical aspects (part A)}

Here we first  define the cross-sections, albedo, scattering
matrix and other
{\it basic optical characteristics of non-spherical
scatterers} and compare them with those for spherical particles.
The idea is to supplement and compress
information given in other sources of such definitions
(see  Mishchenko et al., \cite{Mi00}, \cite{mishetal02})

The {\it parameters of non-spherical scatterers}
(size, shape, etc.)
are briefly
discussed in the next subsection and the light scattering
models popular in astronomy (spheres, cylinders, spheroids, aggregates)
are listed in there.

To realize any scattering model, i.e. to obtain its optical properties,
one should select a proper exact or approximate light scattering method.
 We make an {\it overview of various light scattering methods} and
cite several other useful reviews of this kind.
 A special very detailed review of the methods was prepared
by N.G.~Khlebtsov (see Sect.~3.1 for more details)

The optical properties of cosmic dust analogues can be also measured
in laboratory. References to two available
WWW {\it databases of such measurements}
are presented.

The last subsection is devoted to {\it bibliography on light scattering
theory and its various applications} which is considered
in detail in the Sect.~3.2.

\subsection{DOP subsections on application aspects (part B)}

Here we start with definitions of {\it optical
constants (refractive indices or dielectric functions) of materials}
and link the Jena-St.~Petersburg Database of Optical
Constants for astronomy (JPDOC) developed earlier.
This database described by Henning et al.~(1999) and
J\"ager et al.~(2003)
consists of original laboratory data, references and
links to resources related to measurements and calculations
of the optical constants for materials important for
astronomical applications.
More information on this part is presented in Sect.~3.3.

The next subsection contains a set of {\it light scattering codes}
developed by us and used in astronomical applications.
Links to other collections of such codes are also given.

To test a code not used before one often needs some
well checked results. The optical characteristics
of spheroidal particles of different size, shape, and refractive index
calculated by several methods are given in our
{\it benchmark} subsection.

An important part of the database is the
{\it graphic library of the optical properties} of different
scatterer models --- homogeneous and core-mantle spheres,
homogeneous and core-mantle cylinders, homogeneous
spheroids.
In another part of this subsection
the properties of dust grains which can be estimated
from the observations of
the interstellar extinction and polarization
are discussed (see Sect.~3.4 for more details).

The DOP also includes description and computer realization
of a {\it self-training algorithm to predict the optical properties}
of fractal-like aggregates. As this tool was not yet properly outlined
in the literature, Sect.~3.5 is devoted to its brief description.

\subsection{DOP subsections with related resources (part C)}

The optical properties are very often used
in {\it radiative transfer} (RT) calculations.
Therefore, we concern this point as well as
list the RT methods and
the majority of the RT codes used in astronomy.

The last subsection contains links to different
{\it related resources} (like other databases of light scattering
codes, electronic newsletters, etc.)
which might be of some use in the DOP context.

\section{A more detailed description of some parts of the database}

\subsection{Review of light scattering methods for non-spherical particles (A3)}

The review consists of two parts --- an
bibliographic overview of history and development
of several basic methods and a comparison of all available
approaches to solution of the light scattering problem for
non-spherical particles.

The overview made by N.G.~Khlebtsov
concerns both the exact (separation of variables,
point matching, integral equations, coupled dipoles, T-matrix)
and approximate [Rayleigh, Rayleigh--Gans--Debye (RGD) and its generalizations,
anomalous diffraction, eikonal, geometrical optics (GO), perturbation, etc.]
methods.
In this review the history of development of the methods is traced in very
careful manner. The works on the applicability range and various
applications of the methods are referred as well, and
all together there are over 500 cited papers.

The comparison  of methods is made in form of a table for 11 main approaches:\\
--- separation of variables method, \\
--- discretized Mie formalism, \\
--- finite element method,\\
--- finite difference method,\\
--- point matching method,\\
--- generalized multipole technique,\\
--- extended boundary conditions method,\\
--- method of moments,\\
--- coupled dipoles method,\\
--- Fredholm integral equation method,\\
--- ray tracing/Monte Carlo method.

These approaches are classified and their ability to treat
particles of different shape and structure as well as
anisotropic and chiral scatterers and
ensembles of particles is considered.
 For each method, references to the ansatz and some recent reviews
are given.
Comments to the table show the relationships and
differences of the approaches, briefly concern their classifications
and features, sources of available codes, and refer to
works on comparison of the methods and other reviews.

We have also started the numerical investigation of
applicability ranges of most popular methods
using the spheroidal model of scatterers.
An example of the results is given in Fig.~1 (see Il'in et al.,~\cite{Il02}
for more details).
\begin{figure}\bc
\resizebox{\hsize}{!}{\includegraphics{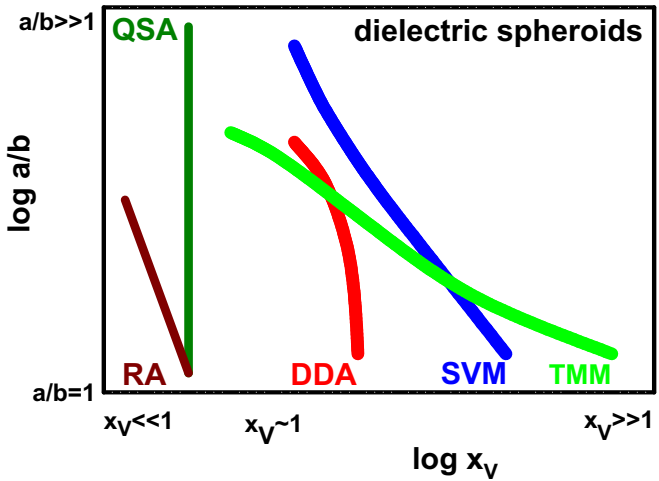}}
\caption{Applicability ranges (domains left to the curves)
of different methods in the case of dielectric prolate spheroids
($a/b$ is the semiaxis ratio,
$x_{V}$ the size parameter of equivolume sphere).
Calculations were made with codes based on
the Rayleigh (RA) and quasistatic (QSA) approximations,
the discrete dipole (DDA), separation of variables (SVM)
and T-matrix (TMM) methods.
}\label{Fig.1}
\ec\end{figure}

\subsection{Bibliography on light scattering and its applications (A5)}

Besides the review, the DOP also includes
a special searchable bibliographic database on light scattering theory
and its various applications.
This bibliography being collected for over 15 years by V.A.~Babenko
consists of the following parts:\\
--- {\it spherical homogeneous particles} --- methodological questions, functions,
   calculations, internal fields, resonances, near field, etc. \\
--- {\it inhomogeneous spheres} --- layered, with radial and chaotic internal
   inhomogeneities, with inclusions, resonances, etc. \\
--- {\it approximations for homogeneous spheres} --- GO, Rayleigh, RGD, eikonal,
   van de Hulst (anomalous diffraction),  and so on approximations; \\
--- {\it complicated problems} --- beam scattering, scattering by charged spheres,
   by rough, anisotropic and chiral particles, etc.; \\
--- {\it cylindrical particles} (see below); \\
--- {\it non-spherical objects} --- works on different approaches and methods; \\
--- {\it specific scattering media} --- interstellar particles, marine particles,
    metallic zoles, powders, etc.; \\
--- {\it carbon}; \\
--- {\it ice}; \\
--- {\it water}; \\
--- {\it agglomerates, fractals}; \\
--- {\it non-linear and mechanical effects in aerosols}; \\
--- {\it vaporization and explosion of particles}; \\
--- {\it inverse problems}; \\
--- {\it laboratory experiments}; \\
--- {\it laser remote sensing}; \\
--- {\it radiative transfer}; \\
--- etc.

To demonstrate the wideness of the questions considered
we shall exemplify a full list of subsections for the section on
cylindrical particles:  \\
--- circular infinite cylinders; \\
--- oblique incidence and orientation averaging; \\
--- multi-layered cylinders; \\
--- GO approximation (including hexagonal cylinders); \\
--- cylinders with smooth radial inhomogeneity; \\
--- gyrotropic, gyroelectric and anisotropic cylinders; \\
--- internal and near fields; \\
--- experiments on cylinders; \\
--- morphological resonances; \\
--- van de Hulst approximation for cylinders; \\
--- Rayleigh and RGD approximations; \\
--- Hart-Montroll and S-approximations; \\
--- non-circular infinite cylinders; \\
--- circular cylinder of finite length and discs.

The total number of included references exceeds 8000
and each subsection contains several dozens of references.

\subsection{Optical constants for cosmic dust analogues (B1)}

\begin{table*}[htb]
\centering
\caption[]{Jena data currently available from the JPDOC}\label{Tab1}
\begin{tabular}{lllcl}
\hline
\noalign{\smallskip}
Compound  & Composition      & State & Low temperature  & Spectral regions \\
 &      &  &  measurements  &  \\
\noalign{\smallskip}
\hline
\noalign{\smallskip}
silicates & (Mg,Fe)SiO$_3$, (Mg,Fe)$_2$SiO$_4$  & glassy & yes  & ultraviolet -- infrared  \\
         & MgSiO$_3$, (Mg,Fe)$_2$SiO$_4$   & crystalline & yes  & infrared          \\
         & MgSi$_x$O$_y$         & amorphous & yes & ultraviolet -- infrared   \\
         & (Ca,Al,Mg,Fe)Si$_x$O$_y$ & amorphous & no & infrared          \\
sulfides & (Mg,Fe)S              & crystalline & yes & infrared          \\
         & SiS$_2$               & crystalline & no   & infrared          \\
oxides   & (Mg,Fe)O              & crystalline & yes  & ultraviolet -- infrared   \\
         & Al$_2$O$_3$           & amorphous & no & infrared          \\
         & (Mg,Al)O$_x$          & crystalline & no & infrared          \\
carbon   & a-C:H                 & amorphous & no  & ultraviolet -- infrared   \\
\hline
\end{tabular}
\end{table*}

This part of the DOP includes some notes introducing the
optical constants required by astronomy, and
links the optical constants database (JPDOC).
The JPDOC consists of two main parts.
One is the bibliography on optical constants of
astronomically interesting materials:\\
--- amorphous/glassy/crystalline silicates of different kinds; \\
--- silicon, SiO, crystalline/fused SiO$_2$;\\
--- metals: Fe, Mg, and others; \\
--- oxides: FeO, Fe$_2$O$_3$, Fe$_3$O$_4$, MgO, Al$_2$O$_3$, MgAl$_2$O$_4$; \\
--- sulfides: FeS, MgS, SiS$_2$; \\
--- carbides: SiC, FeC, TiC; \\
--- carbonaceous species: diamonds, graphite, coals, kerogens, HAC,
   glassy/amorphous carbon, PAHs and so on; \\
--- organics: tholin, ``organic refractory", etc. \\
--- ices: H$_2$O, CO, CO$_2$, NH$_3$, HCN, etc. and their mixtures; \\
--- FeSi, CaCO$_3$ and some other materials.\\
We tried to include all papers published
since the beginning of the previous century
(all together over 700 references).

Another part is a set of data specially measured
in the Jena laboratory for astronomical purposes.
Note that although various terrestrial analogues of
cosmic solids have been studied in chemical and
physical laboratories, many of these experiments
neither took into account the specifics of cosmic
dust materials (composition, lattice structure,
etc.) and conditions (low temperature, processing,
etc.), nor covered the wavelength
intervals of the current astrophysical interest.

A summary of the data obtained in Jena and made
available is given in Table~\ref{Tab1}; data examples
are presented in Fig.~2 (see J\"ager et al., 2003
for more examples).

\begin{figure}\bc
\resizebox{\hsize}{!}{\includegraphics{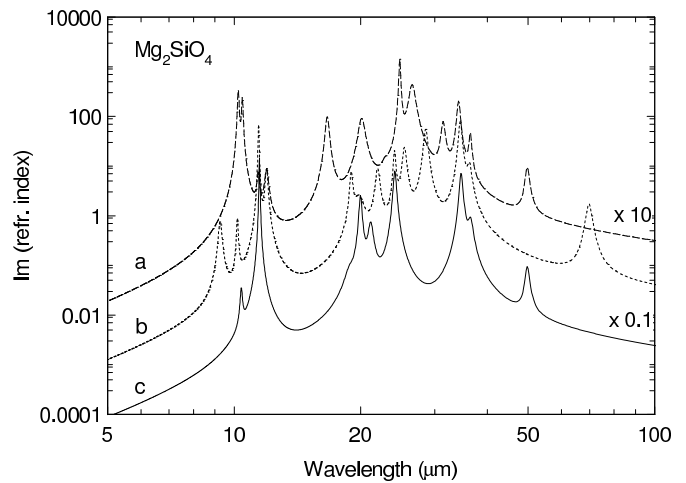}}
\caption{ Imaginary part of the refractive index for
crystalline forsterite (Mg$_2$SiO$_4$) in the three
different crystallographic directions.
}\label{Fig.2}
\ec\end{figure}

\subsection{Graphic library of optical properties of non-spherical scatterers (B4)}

This DOP section consists of two parts ---
a basic graphic library and
special notes on optical properties of cosmic dust analogues and
their connection with observed manifestations of interstellar dust.

In the basic library we show how different optical characteristics
(extinction, scattering and absorption efficiencies, albedo,
asymmetry factor, etc.)
vary with scatterer parameters: the real ($n$) and imaginary ($k$) parts
of the refractive index, size
(the refractive index is kept independent of radiation wavelength),
shape, structure.
So far we have considered homogeneous and core-mantle spheres
and infinite cylinders, and homogeneous spheroids, but are planning
to involve more scatterer models.
The figures in the DOP are given for weakly absorbing
($n = 1.3 - 1.7$, $k = 0 - 0.3$) and highly absorbing
($n = 1.5 - 3.5$, $k = 0.5 - 2.5$) particles
(see Fig.~3 for examples). These refractive
indices are typical of astronomically interesting
materials in visual.

\begin{figure}\bc
\vspace{10cm}
\resizebox{\hsize}{!}{\includegraphics{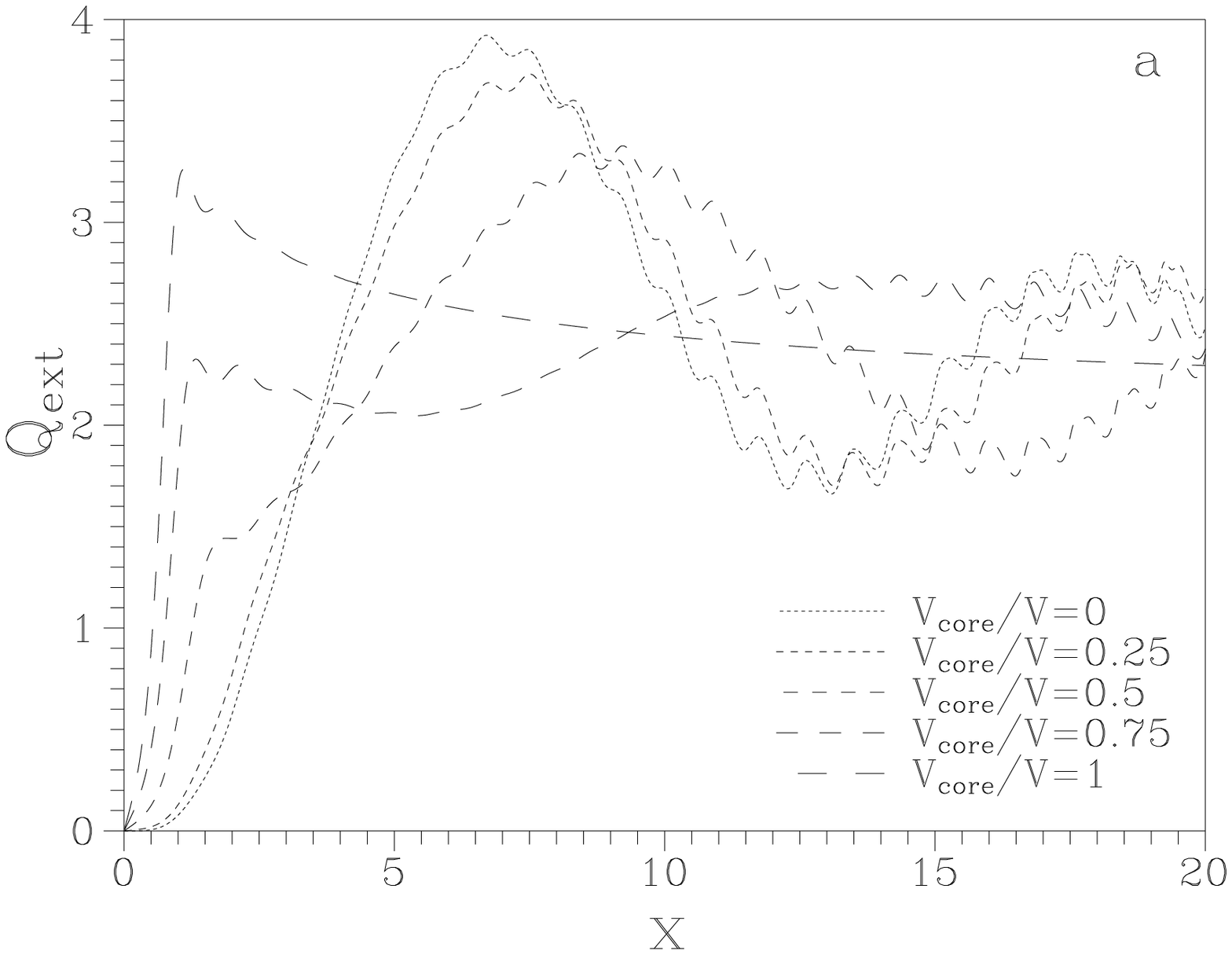}}
\resizebox{\hsize}{!}{\includegraphics{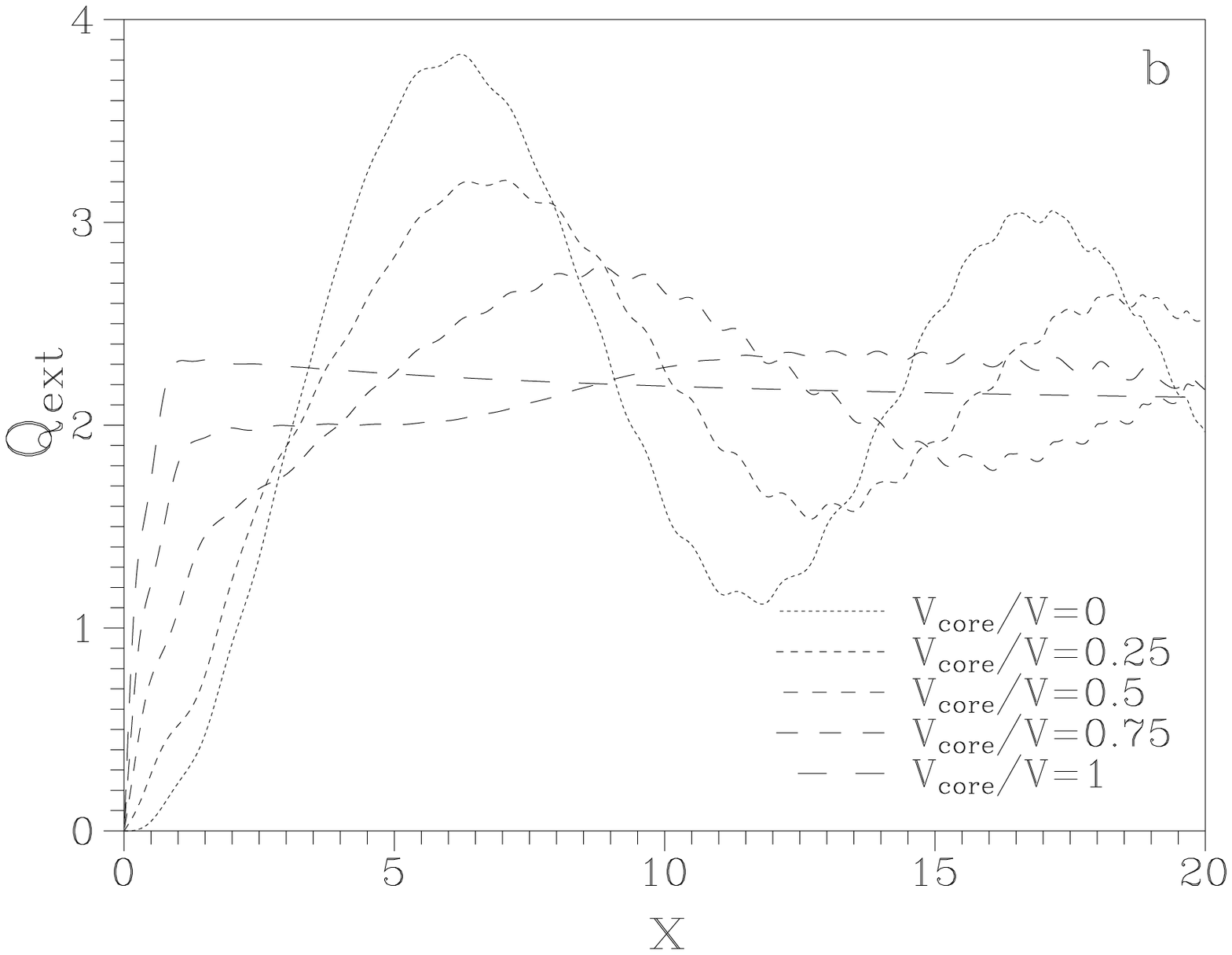}}
\caption{The extinction factors for
core-mantle spheres (a) and infinite cylinders (b)
of different core size (volume).
The refractive index of the core is $m = 2.5+1.5i$ and
that of the mantle $m = 1.3+0.0i$
(the normal incidence of light on the cylinders).
}\label{Fig.3}
\ec\end{figure}

The special notes are the chapters of the review of Voshchinnikov~(\cite{Vo02})
which cover the following questions of observed phenomena modelling ---
extinction efficiencies: general behaviour and deviations,
wavelength dependence, the 2175 A feature,
absolute extinction and abundances, etc.;
\ polarization efficiencies: size/shape/orientation effects,
linear polarization: wavelength dependence,
circular polarization: change of sign and so on.
Some more chapters will be included in the near future.

\subsection{Self-training algorithm for fractal-like aggregates (B5)}

This part of the DOP was developed in Kharkov University
by V.P.~Tishkovets, S.A.~Beletsky and P.V.~Litvinov.
Its idea is as follows --- as calculations
of the optical properties of subparticle aggregates
need a huge amount of processor time, it is worthwhile
to construct a self-training algorithm that could
learn to predict the properties analyzing a limited
set of these data.

To reduce the number of free parameters, fractal-like
aggregates of equal size subparticles were selected.
The structure of the aggregates can be characterized
by two parameters: the fractal dimension $D$
and the prefactor constant $\rho$ (see, e.g., Feder, 1988).
Other parameters are the number of subparticles $N$,
the size parameter $x$ and the real and imaginary
parts of the refractive index of subparticles.

The algorithm was based on an artificial neural network
named 'multi-layered perceptron'
(see, e.g., Miller, 1990 for more details).
A model of perceptron with 6 input and 30 output neurons
and two layers with 30 neuron in each between the input and output
ones was used.
The algorithm was realized in the form of Pascal and Fortran codes.
The region of its training was $n = 1.4 - 1.7$; $k = 0.001 - 0.1$;
$x = 1.5$; $D = 3$; $\rho = 8$; $N \le 50$.
Accurate calculations were done by the Mackowski and Mishchenko (1996) code.
The results were the expansion coefficients of the scattering
matrix elements in series of the generalized spherical functions
for particles of different $N$, refractive index, etc.
Figure 4 shows the accuracy of the perceptron results.

The codes and data archive as well as documentation
are available via the DOP. Our work with the perceptron
on modelling of the optical properties of interplanetary
dust grains
has demonstrated its advantages due to extremely small size
of the codes and very fast data access.

\begin{figure}\bc
\vspace{6cm}
\resizebox{\hsize}{!}{\includegraphics{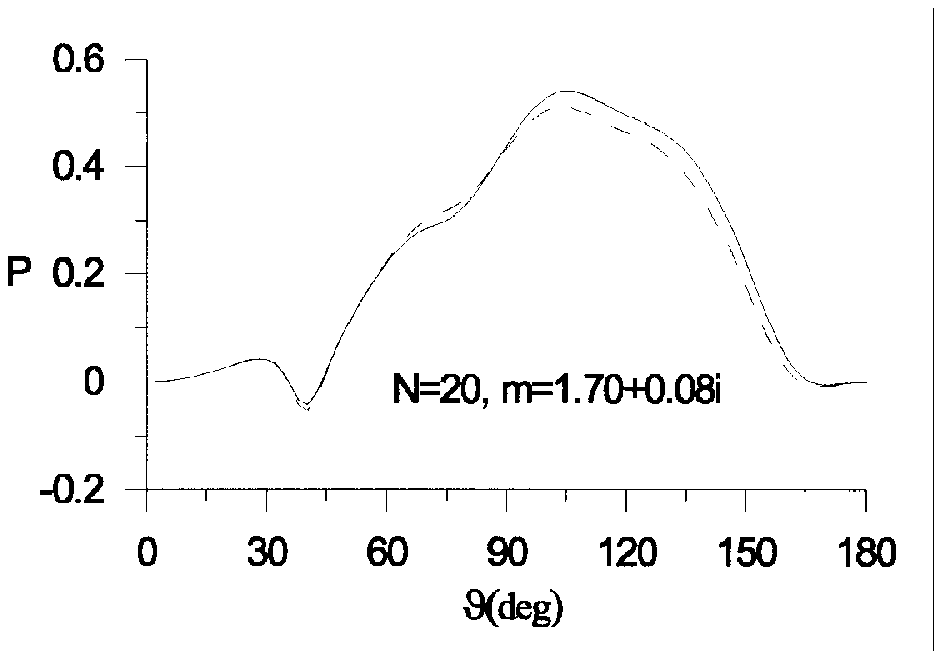}}
\caption{Linear polarization degree $P$
versus the scattering angle $\theta$. Solid curve
corresponds to the accurate values, dotted one to values
given by the perceptron.
The number of subparticles in an aggregate $N = 20$,
$m = 1.7+0.08i$.
}\label{Fig.4}
\ec\end{figure}

\section{Concluding remarks}

We have introduced a database of the optical properties
of cosmic dust analogues.
 It is designed for student and scientists to help
in applications of non-spherical scatterer models
in astronomy but can be useful in other scientific fields as well.

 A further development of the database is planed in near future.
It will include a graphics interface for the JPDOC,
a special tool for on-line calculations of different
optical properties of non-spherical layered particles,
more chapters from Voshchinnikov's~(\cite{Vo02}) review
and its continuation, and so on.

\acknowledgements{
The authors thank V.G.~Farafonov, S.I.~Grachev and D.I.~Nagirner
for useful discussions and comments.
The work was partly supported by
the INTAS (Open Call 1999 grant N~652),
the Russian Ministry of High Education,
the Russian federal program Astronomy
and that for prominent scientific schools.
}



\end{document}